\documentclass[12pt,onecolumn]{article} 
\usepackage{xcolor}
\usepackage{graphicx}
\usepackage{url,hyperref}
\usepackage{amsmath,amssymb,amsfonts,upref,endnotes}
\usepackage{amsthm}
\usepackage{marginnote}
\usepackage{soul}
\usepackage{color}

\begin{document}

\title{The response of a bistable energy harvester to different excitations: the harvesting efficiency and links with stochastic and vibrational resonances}

\author{
I. A. Khovanov \\
School of Engineering, University of Warwick, Coventry, CV4 7AL, UK \\
i.khovanov@warwick.ac.uk
}

\maketitle
\thispagestyle{empty}

\begin{abstract}
	Energy harvesting of ambient vibrations using a combination of a mechanical structure (oscillator) and an electrical transducer has become a valuable technique for powering small wireless sensors. Bistable mechanical oscillators have recently attracted the attention of researchers as they can be used to harvest energy within a wider band of frequencies. In this manuscript, the response of a bistable harvester to different forms of ambient vibration is analysed. In particular, harmonic noise, which has a narrow spectrum, similarly to harmonic signals, yet is stochastic, like broad-spectrum white noise, is considered. Links between bistable harvester responses and stochastic and vibrational resonance are explored.
\end{abstract}
\section{Introduction}

Powering sensors and wireless communication using energy harvesting has become a crucial technological development leading to a variety of applications \cite{Priya:09}. 
A typical energy harvester, which converts ambient vibrations to electrical power, consists of a mechanical structure and a converter \cite{Priya:09,Anton:07}. Designing a harvester with optimal power output is a complex problem which continues to attract the attention of engineers: the literature reveals a complex interplay between the properties of ambient vibration, the particular design of the mechanical structure and the type of converter used \cite{Gholikhani:20}. 

A variety of mechanical structures and types of converter have been suggested in the literature \cite{Priya:09,Anton:07,Gholikhani:20}. However, the properties of ambient vibrations have attracted significantly less attention. Typically, ambient vibrations are assumed to be harmonic, i.e. in the form of a sine wave signal. In cases where the strongest contribution comes from periodic vibrations, for example from a rotor with a fixed frequency, the stochasticity of the vibrations is negligible. A harmonic signal is thus a valid representation of these ambient vibrations. In many other cases, such as in a car with many different modes and sources of vibration, however, representing vibrations via a harmonic signal becomes too simplistic. A better representation of these ambient vibrations is white noise \cite{Cottone:09,Litak:10,Khovanova:11,Khovanova:14}.

A third model for ambient vibrations as a narrowband signal in the form of harmonic noise has also been considered \cite{Khovanova:14}. Unlike harmonic oscillations, harmonic noise is stochastic; however, its power is concentrated in a much narrower band of frequencies than that of white noise \cite{Schimansky:90,Dykman:91,Khovanov:17}. In this sense, harmonic noise is intermediate between a harmonic signal and white noise \cite{Khovanov:17}. Harmonic noise describes ambient vibrations well in civil structures which have a dominant vibration mode excited by a variety of external factors\cite{Penzien:93,Lin:04}; a harvester attached to such structures is therefore excited by stochastic narrowband perturbations. One example is a bridge with passing cars and trucks.

If the harvester is adequately described by a linear system, the properties of ambient vibrations are not a significant factor in the design. In this case, matching the system's frequency response to the spectrum of vibrations is an efficient strategy \cite{Beeby:06,Priya:09}. However, linear behaviour is usually observed only within a limited range of vibrational powers, and the inherent nonlinearity of the harvester makes such a matching strategy ineffective. In the case of a strongly nonlinear response, other approaches are used, for example increasing the bandwidth of the harvesting vibrations\cite{Erturk:09}. An example of such nonlinear design is a harvester with a mechanical part with a multistable configuration \cite{Cottone:09,Litak:10,Kim:14,Zhou:18}. In particular, a bistable (double-well) configuration was suggested as an effective design in the case of broadband vibrations \cite{Cottone:09,Litak:10,Khovanova:11}. Additionally, it was suggested that the presence of  stochastic resonance (SR) \cite{McInnes:08} and vibrational resonance (VR) could be used to find an optimal bistable configuration for a harvester excited by a mixture of two signals \cite{Coccolo:14}. These phenomena are widely considered to be a way to enhance a system's response to a low frequency signal by adding either a stochastic component (in SR) \cite{Gammaitoni:98,Anishchenko:99}, or a high-frequency harmonic signal (in VR) \cite{Landa:00}.

The mechanisms of these two phenomena are widely discussed, albeit in a variety of different and often contradictory interpretations: the literature is vast and growing, and there is no clear consensus. Nevertheless, a typical feature of SR and VR is a transition from intra-well to inter-well dynamics; here ``well'' refers to the area around an equilibrium state. In a typical setting, a bistable system is excited by a weak low-frequency harmonic signal, which alone is unable to induce switching between wells. An additional signal, which is noise in SR and a high-frequency harmonic signal in VR, is able to induce the transition from intra- to inter-well dynamics. This transition affects the system's response, and results in an ``optimal'' response when the additional signal has a particular magnitude. Optimality occurs when certain variables describing the system's response are extremised. Commonly used indicators are the coherence factor in VR and amplification factor in SR; these both characterise the height of the peak in the system's frequency response at the frequency of the input harmonic signal and are therefore equivalent. The phenomena typically appear in an overdamped bistable system which is characterised by just one relaxation time scale. The period of the harmonic signal should be shorter than the relaxation scale of the system for either phenomenon to manifest strongly.
 
A typical harvester configuration requires the mechanical part to be underdamped. This requirement adds a time scale linked to the resonance frequency which must be taken into account when considering SR and VR in a bistable harvester. Another distinct feature of a harvester is that the velocity of the mechanical part defines its efficiency, whereas it is the coordinate (position) which is typically used for characterising SR and VR \cite{Coccolo:14}. However, the relationship between underdamped harvester dynamics, SR and VR indicators, and harvester efficiency remains obscure. In order to untangle this relationship, in this work,  the responses of a bistable harvester to the three types of signal discussed above, along with the links between efficiency and SR/VR indicators, are considered.

\begin{figure}[!h]
\includegraphics[width=2.25in]{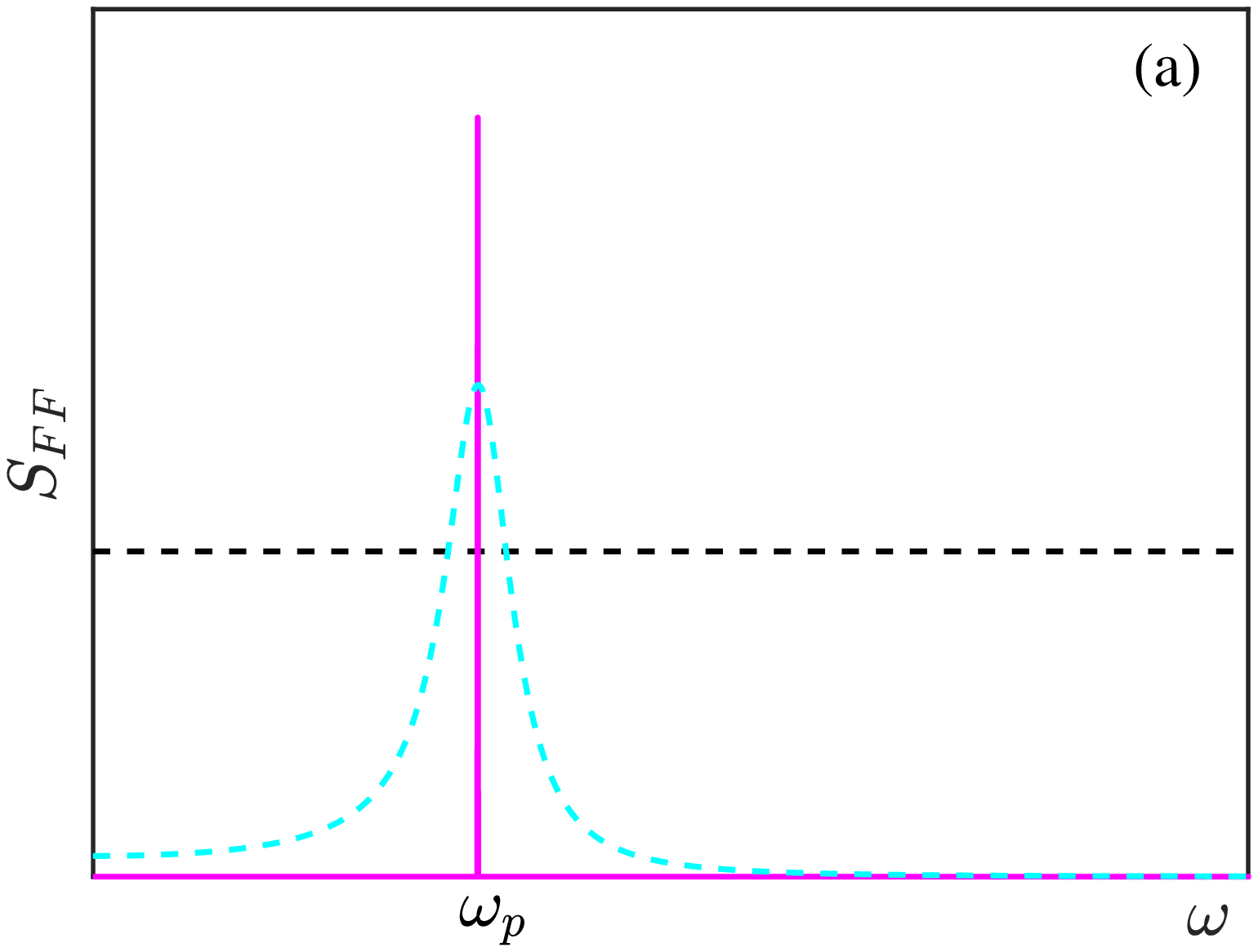}~~\includegraphics[width=2.25in]{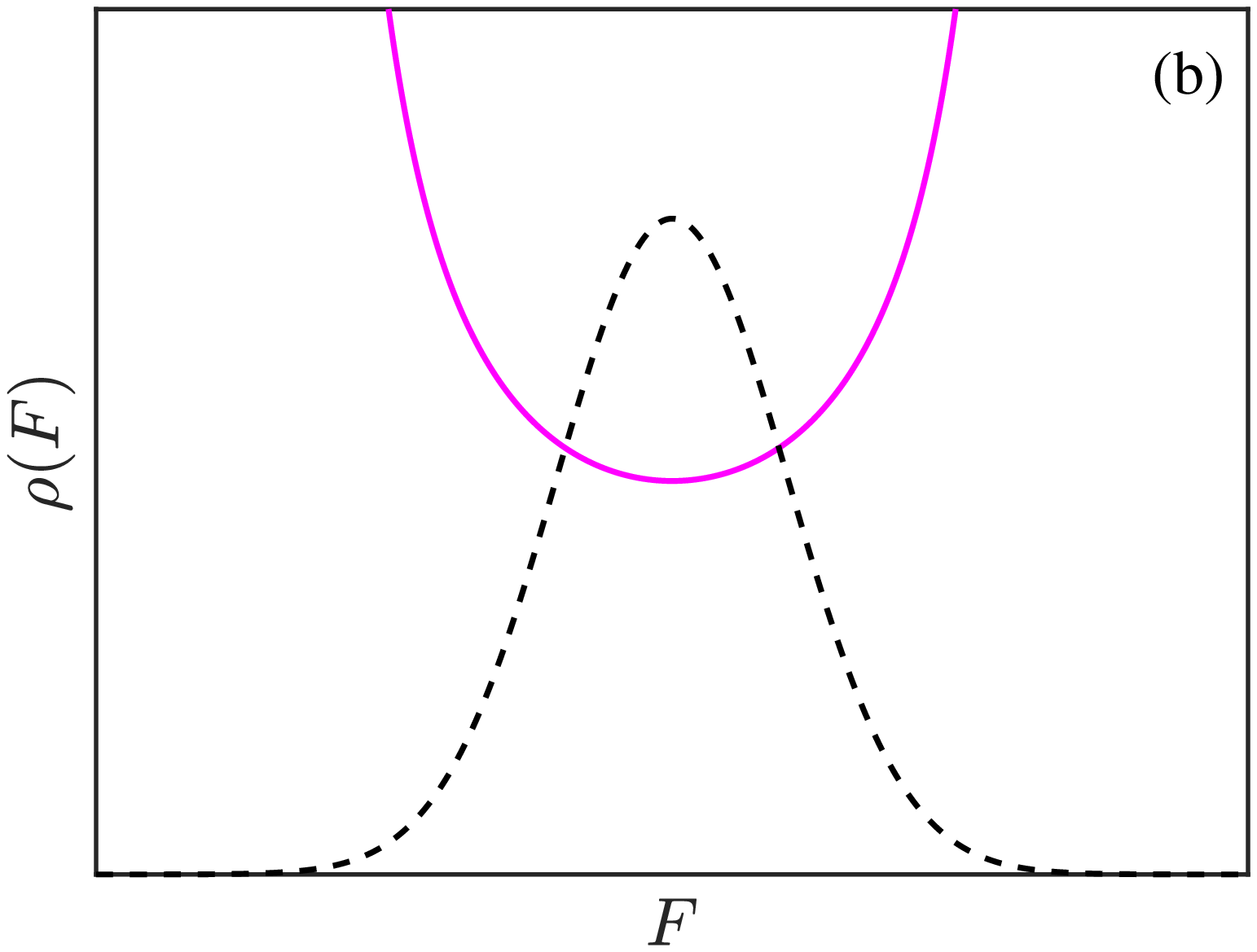} 
	\caption{(a) Examples of spectral shapes, $S_{FF}$, for different forms of ambient vibrations, $F(t)$, are shown.   Solid magenta line,  dashed cyan line and dashed black line correspond to the harmonic signal, harmonic noise and white noise respectively. (b) Examples of probability density distribution, $p(F)$, for different forms of $F(t)$ are shown. Solid magenta line corresponds to the limited distribution of the harmonic signal. Dashed black line corresponds to Gaussian distribution of harmonic and white noises. }
	\label{fig00}
\end{figure}

	Spectra and distributions of these three signals are shown in Fig.~\ref{fig00}. The spectral content of the harmonic noise and harmonic signal is localised around a particular peak frequency, $\omega_p$, and the white noise has a flat spectrum. The distributions of both noises are Gaussian with wide tails, and the distribution of the harmonic signal is limited. Thus, the harmonic noise is close to the harmonic signal in the spectral domain, but similarly to white noise has the Gaussian distribution. Comparison the responses to these signals and their combinations allows us, therefore, to clarify which properties of these signals define the manifestations of SR and VR phenomena and the harvester efficiency. The consideration of different peak frequencies, $\omega_p$, for the harmonic noise and the harmonic signal help to unravel the poorly understood role of the underdamped dynamics of the harvester.  
	
	For the task of engineering design of the bistable harvester, it is important to know the height of the barrier between the states with respect to the magnitude of the external excitation. A ratio between the height and the magnitude defines the closeness of an operating point of the bistable harvester to the point of the transition from intra- to inter-well dynamics. Despite a larger volume of the literature, the links between the efficiency of energy harvesting and SR/VR indicators with respect to the point of the transition are unclear. In this work, therefore, this transition point is considering as a design parameter for an efficient harvester. 
	
	The paper has the following structure. 
	In section 2, the system under consideration is described, together with the quantities used to characterise the system response. Details of numerical simulations are also given. In section 3, the response of a bistable harvester to harmonic noise is compared with the responses to a harmonic signal and white noise. This comparison allows us to clarify the role of the underdamped dynamics of the bistable harvester and the properties of ambient vibration on harvester's efficiency. In particular, the comparison unravels links between the extremal values of the system's characteristics and the transition from intra- to inter-well dynamics. In section 4, SR and VR are analysed, with the additional signal being either white noise, harmonic noise or a high-frequency harmonic signal. The consideration of different peak frequencies for the harmonic noise and the harmonic signal allows us to clarify the role of the underdamped dynamics of the harvester. Note that the combinations of the harmonic signal and white noise for SR and low- and high-frequency harmonic signals for VR were discussed previously in the literature. However, in the best of author's knowledge, the combination of the harmonic signal and high-frequency harmonic noise has not been considered before.  The comparison of the system response to these different combinations elucidates the influence of the properties of the additional signal on the efficiency of the harvester, and hence the links to SR and VR. The results are summarised in Section 5.

\section{System and characterisation of its response}     

A model of a bistable harvester widely used in literature \cite{Litak:10} consists of a Duffing oscillator, which represents the mechanical part, coupled with an additional first order equation, which corresponds to the electrical part:  
\begin{align}\label{eqsys}
	\begin{split}
	&	\ddot{x} + \alpha \dot{x} +0.5 (x^3-x)+\psi v = F(t)      \ , \\
&		\dot{v}+ \lambda v= - \kappa \dot{x} \  .
	\end{split}
\end{align}
Coordinate $x$ and velocity $\dot{x}$ define the dynamics of the mechanical part of the harvester --- typically a beam. A piezo-electric component, typically consisting of a piezo-electric strip or a pack of strips connected to an electrical circuit, is described by voltage $v$.  In model (\ref{eqsys}), the damping factor $\alpha=1$ represents both internal damping of the mechanical beam and additional damping in a piezo-electric strip attached to the beam. The shape of the bistable potential is selected so that the natural frequency of the stable state is equal to 1. Parameter $\psi=0.05$ represents the small back action of the piezo-electric strip. The transformation of mechanical vibrations to voltage is described by parameter $\kappa=0.5$. The relaxation time of the electrical part is defined by parameter $\lambda=1$. Since the back action defined by $\psi$ is small, the dynamics of voltage $v$ and velocity $\dot{x}$ are strongly correlated. Signal $F(t)$ corresponds to ambient vibrations. 
Note that since the harvester consists of three independent components: a mechanical part (beam), piezo-electric strips and an electrical circuit (load), the parameters of the harvester can vary in a broad range.

The efficiency of energy transfer can be characterised by the power ratio $e=P_{out}/P_{in}$ of the input $P_{in}$ and output $P_{out}$ powers \cite{Khovanova:11,Coccolo:14}. In the following sections, two ratios $e_x$ and $e_v$ are considered, with output powers defined by expressions:   
\begin{align}\label{eqPx}
			P_{out}=P_x=\frac{1}{\tau}\int_{t_r}^{t_r+\tau} ( x(t) -\mu_x ) ^2 dt       \ , 	\ \mu_x=\frac{1}{\tau}\int_{t_r}^{t_r+\tau} x(t)  dt \ ,
\end{align}
and 
\begin{align}\label{eqPv}
	P_{out}=P_v=\frac{1}{\tau}\int_{t_r}^{t_r+\tau} v^2(t) dt       \ , 	
\end{align}
respectively. In these expressions, $t_r$ is the transient time and $\tau$ is the estimation interval. Note that the ratio $e_v$ reflects the efficiency of energy harvesting whereas $e_x$ does not, instead describing the dynamics of coordinate $x$ in (\ref{eqsys}). 

To characterise SR and VR, the coherence factors $\gamma_x$ and $\gamma_v$ were analysed. The factors are defined by the following expressions \cite{Coccolo:14}:
\begin{align}\label{eqgamma}
	\gamma = \frac{2\sqrt{C_s^2+C_c^2}}{A} \ , \ \  	
	C_s=\frac{1}{\tau}\int_{t_r}^{t_r+\tau} s(t) \sin(\omega t) dt       \ , \ \ \    C_c=\frac{1}{\tau}\int_{t_r}^{t_r+\tau} s(t) \cos(\omega t) dt       \ , 	
\end{align}
where $s(t)\equiv x(t)$ is used for $\gamma_x$ and $s(t)\equiv v(t)$ for $\gamma_v$; $A$ and $\omega$ are the amplitude and frequency of the low-frequency harmonic component of the ambient vibrations. The estimation interval $\tau = nT$ is defined as an integer number $n$ of periods $T=2\pi / \omega$ of the harmonic signal.

Model (\ref{eqsys}) was numerically simulated using the Heun method \cite{Mannella:02} for both deterministic and stochastic forms of $F(t)$. The integration step size was selected to be a multiple of period $T$ if a harmonic signal was present in $F(t)$; in this case the step size was also chosen to be smaller than 0.01. For a pure stochastic form of $F(t)$, the step size was equal to 0.01. The initial conditions, estimation interval $\tau$ and transient time $t_r$ were varied depending on the form of $F(t)$.

The transition from intra-well to inter-well dynamics was identified using the mean value $\mu_x$ (see Eq. (\ref{eqPx})). The value of $\mu_x$ is close to zero when the dynamics include both states (wells), and differs from zero if dynamics occur only in the vicinity of one of the two states.

\section{Single input}

Model (\ref{eqsys}) is a strongly nonlinear system which demonstrates different responses depending on the magnitude of external excitation $F(t)$. For weak excitations, the system response is described adequately by the response of a linear oscillator with resonance frequency $\omega_r$. At a moderate excitation magnitude, the response is close to that of an oscillator demonstrating soft nonlinearity. In this case, the system has two distinct symmetric states. Further increase in the magnitude of excitations induces switching back and forth between the two states, making bistability a dominant feature in the response of (\ref{eqsys}). At a very large excitation magnitude, system (\ref{eqsys}) is effectively monostable and shows a response typical of an oscillator demonstrating hard nonlinearity. Such variability in the system response significantly complicates the design of an efficient harvester. 

\begin{figure}[!h]
\includegraphics[width=2.25in]{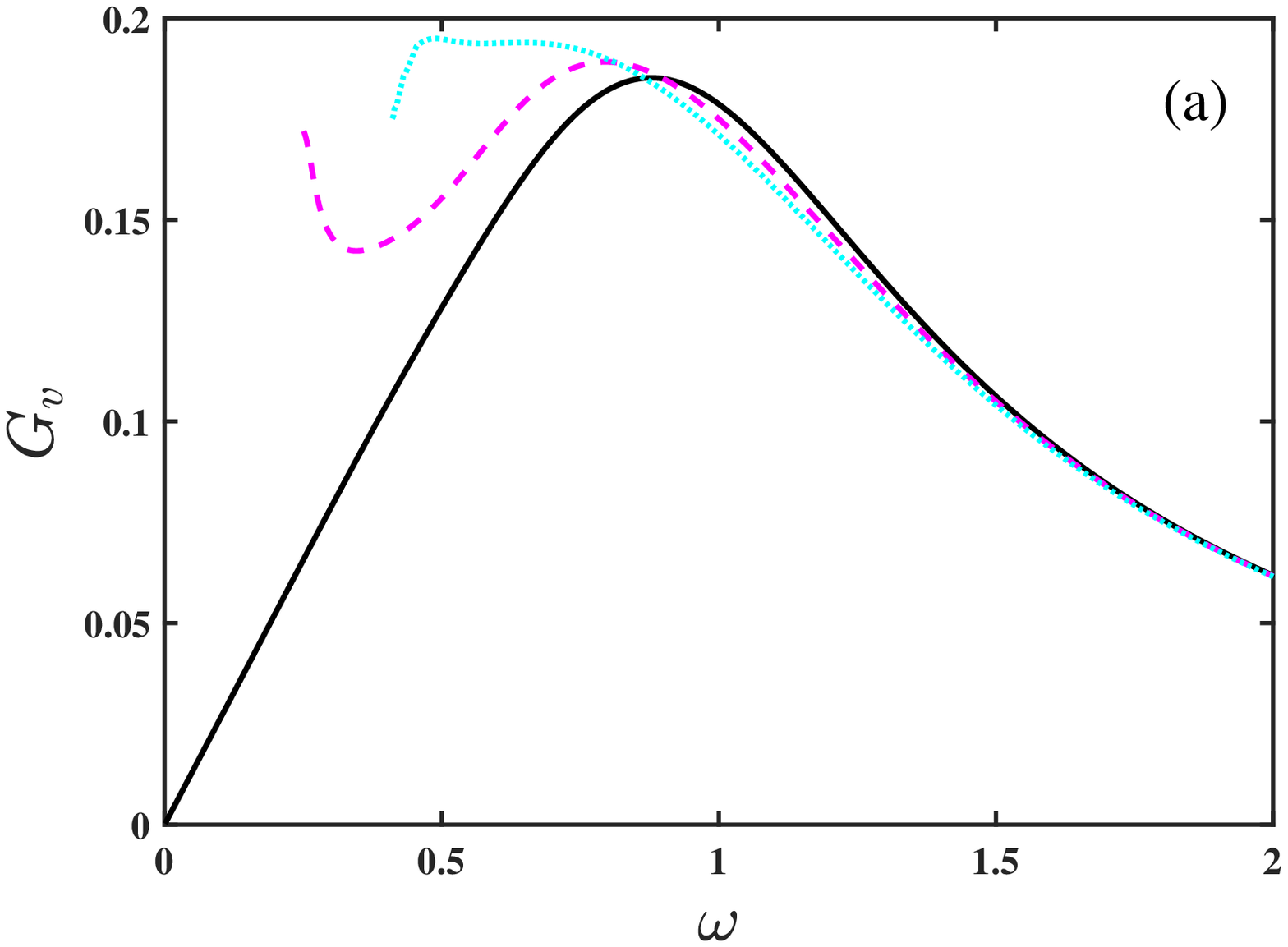}~~\includegraphics[width=2.4in]{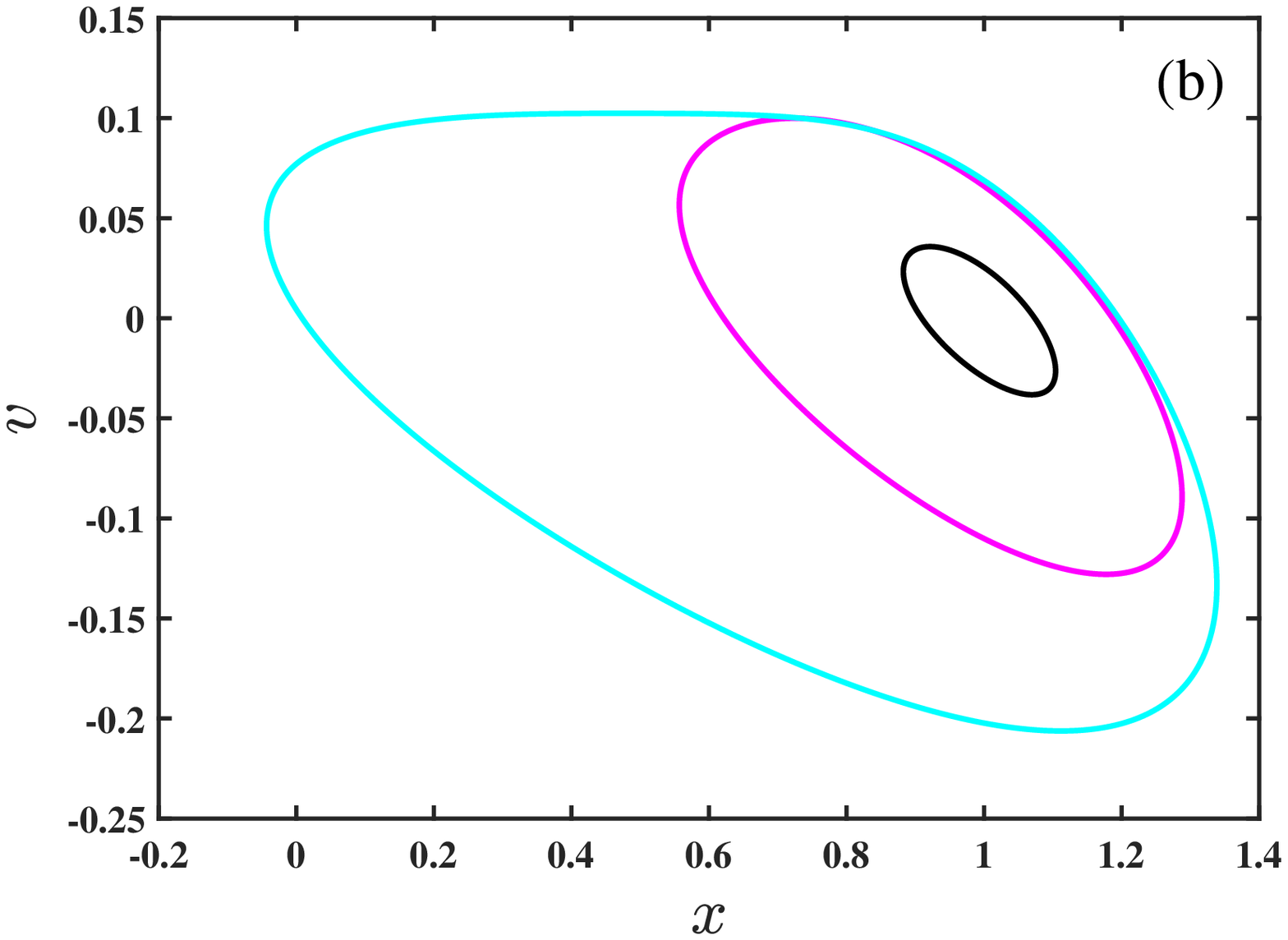} 
\caption{(a) The frequency response of the harvester (\ref{eqsys}) for different values of amplitude $A$: $A=0.1$ (solid black line), $A=0.3$ (dashed magenta line) and $A=0.39$ (dotted cyan line). The quantity $G_v=\sqrt{2 P_v} / A$ corresponds to the gain factor for coordinate $v$ of (\ref{eqsys}); $P_v$ is defined by expression (\ref{eqPv}). The gain factor is shown in the regime of inter-well motion only.  (b) Trajectories of the system  (\ref{eqsys}) are shown on the plane $x$--$v$. The trajectories correspond to maximal gain for each amplitude value used in figure (a). }
	\label{fig0}
\end{figure}

A widely used linear design is based on matching the resonance frequency $\omega_r$ to the frequency corresponding to the strongest peak in the spectrum of ambient vibrations. For system (\ref{eqsys}) linearised about one of the stable equilibrium states, $\omega_r \approx 0.9$. The linear design assumes that the frequency response of system (\ref{eqsys}) does not depend on the amplitude of the harmonic signal. However, the application of the linear design is valid in a limited range of the amplitude. Fig.~\ref{fig0}(a) shows the frequency responses of  (\ref{eqsys}) for several amplitudes. If amplitude $A\lessapprox 0.1$  the frequency response is described well by solid black  line in Fig.~\ref{fig0}(a). However the shape of the frequency response is changed for larger amplitudes, and the response becomes nonlinear, that is it depends on the amplitude value. The soft nonlinearity of intra-well dynamics means that the resonance frequency of the nonlinear response is smaller. The frequency responses shown in Fig.~\ref{fig0}(a) for system (\ref{eqsys}) illustrate this influence of the soft nonlinearity.  Several distinct frequencies should therefore be considered to characterise the response of system (\ref{eqsys}). The three following frequencies were selected: $\omega_1=0.1$, $\omega_2=0.5$ and $\omega_3=0.9$. Frequency $\omega_1$ is typical for observing SR and VR as this frequency is much lower than the relaxation rate of the system. The other two frequencies, $\omega_2$ and $\omega_3$, correspond to the nonlinear (dotted cyan line in Fig.~\ref{fig0}(a)) and linear (black solid line in Fig.~\ref{fig0}(a)) resonance frequencies, respectively.
 
\subsection{Harmonic signal}
\begin{figure}[!h]
	\includegraphics[width=2.25in]{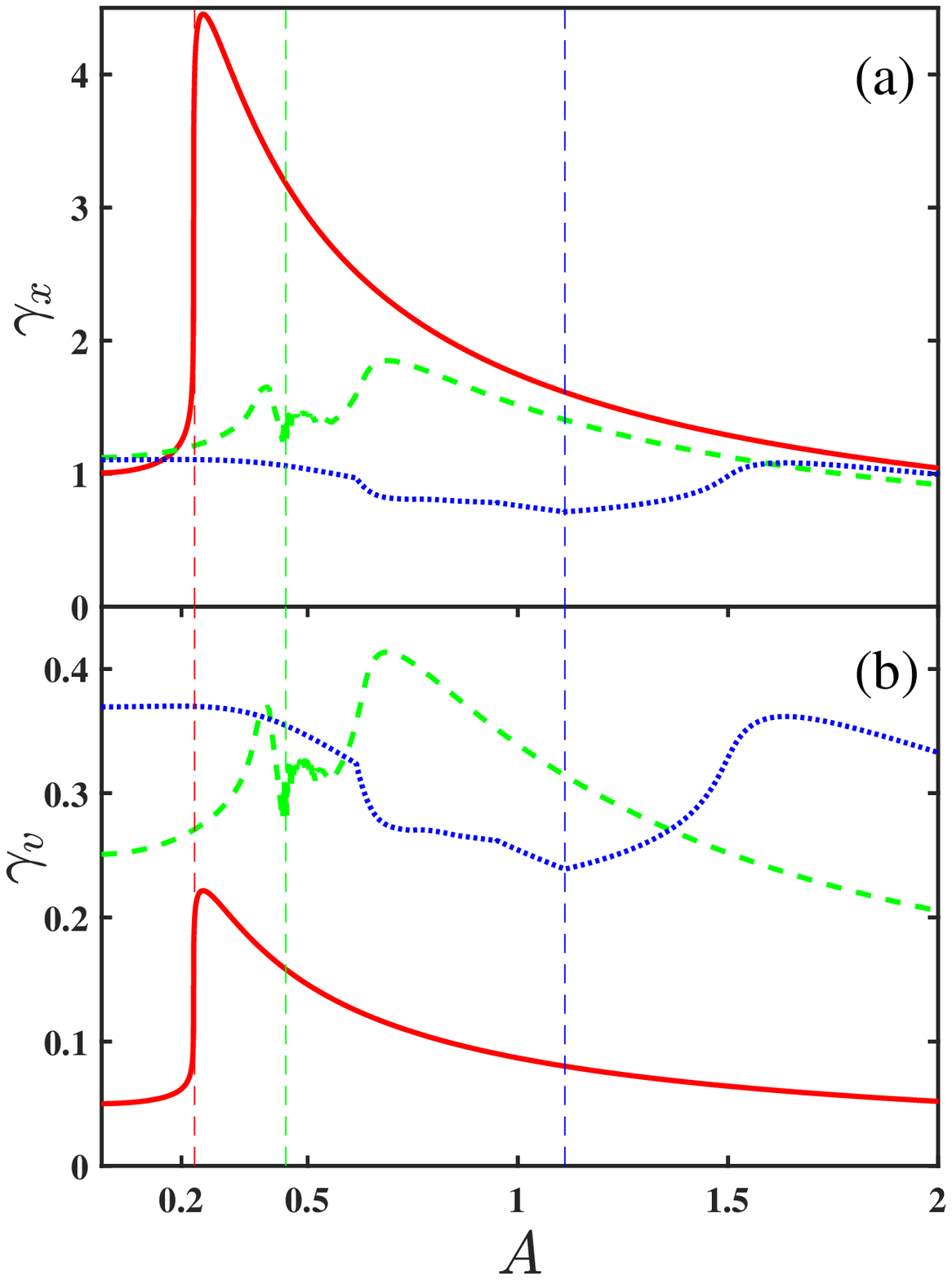}~~\includegraphics[width=2.25in]{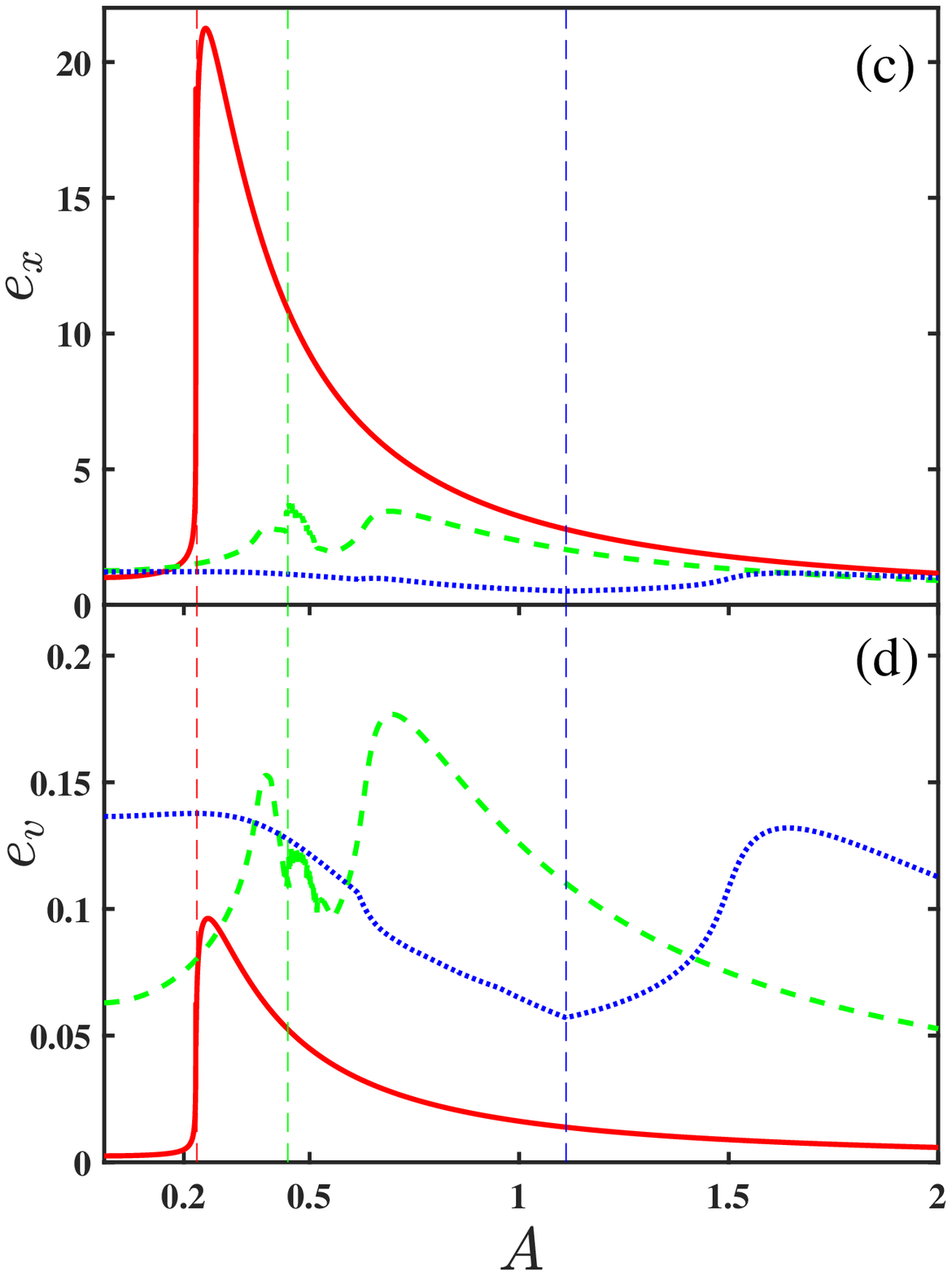}
	\caption{Dependence of coherence factors $\gamma_x$ (a) and $\gamma_v$ (b) and power ratios $e_x$ (c) and $e_v$ (d) on amplitude $A$ of harmonic signal (\ref{eqLFsig}). Solid red, dashed green and dotted blue lines correspond to frequencies $\omega_1=0.1$, $\omega_2=0.5$ and $\omega_3=0.9$ respectively. Vertical dashed lines indicate values of $A$ when the transition to inter-well dynamics occurs.}
\label{fig1}
\end{figure}

Coherence factors and efficiency ratios for ambient vibrations in the form of a harmonic signal
\begin{align}\label{eqLFsig}
	F(t)= A \sin (\omega t)  \  	
\end{align}
are shown in Fig.~\ref{fig1} for the three selected frequencies. The input power for $e_x$ and $e_v$ is given by $P_{in}=A^2 / 2$. Dashed vertical lines indicate the values of amplitude $A$ for which the transition from intra-well to inter-well dynamics occurs. In the case of the lowest frequency $\omega=\omega_1$ (red solid curves in Fig.~\ref{fig1}), this transition dramatically changes the system response: all characteristics demonstrate significant growth after it. Such growth is the primary motivation for the bistable design. After the transition, a further increase in the magnitude of the external excitation leads to a decrease in the characteristics. As a result, there is a relatively narrow range of excitation magnitudes within which the response is optimal. The narrowness of this range presents a challenge for a bistable design since the excitation magnitude generally varies over a wide range. The optimal magnitude range is determined by the rate of switching between states as well as by the relaxation time of the system. The role of intra-well dynamics is insignificant since frequency $\omega_1$ is lower than the resonance frequency of (\ref{eqsys}). 

The behaviour of the characteristics for frequencies $\omega_2$ and $\omega_3$ is dramatically different, however, due to the importance of intra-well dynamics. For $\omega_3$, all characteristics initially decrease with the growth of magnitude $A$ and reach a minimum at the transition amplitude, where the dynamics start to include both states. The bistability plays a destructive role and the characteristics are pessimal (anti-optimal) in the vicinity of the transition point. This result is not surprising since $\omega_3$ was selected due to resonance in the linearised system (\ref{eqsys}). In contrast, nonlinearity leads to deviation from these conditions, and this deviation is strongest at the point where bistability has the greatest influence. Further increase in amplitude $A$ effectively leads to  single-well dynamics. The system response is nonlinear since this single-well state is dominated by the cubic term $x^3$ in (\ref{eqsys}). Therefore, the peak in characteristics around $A\approx 1.65$ corresponds to the nonlinear resonance condition \cite{Blekhman:04}. 

The system's response for frequency $\omega_2$ is more complicated than that in the two previous cases. There are two peaks in the characteristics at $A \approx 0.4$ and $A \approx 0.7$. Both peaks are related to nonlinear resonance in the single-well dynamics. The origin of the peaks is similar to that in the case of $\omega_3=0.9$. However, in contrast to the $\omega_3$ case, the bistability in the case of $\omega_2$ leads to moderate growth in all characteristics at the point of transition to inter-well dynamics. This change is masked by the chaotic dynamics observed in the vicinity of the transition. 

The intra-well dynamics thus significantly affect the influence of bistability on the characteristics considered above. The underdamped dynamics of a bistable harvester should therefore be taken into account when designing it. Note that other states appear at very large magnitudes of $A$; these are not taken into consideration since they are not directly connected to bistability.

\subsection{Stochastic signal}

In this section, two forms of stochastic ambient vibration are considered: white and harmonic noise. Both types of noise have Gaussian (normal) distribution with zero mean and variance (intensity) $\sigma$. The latter determines the input power $P_{in}=\sigma$ for $e_x$ and $e_v$. White noise has a flat spectrum, whereas the spectrum of harmonic noise has a Lorentzian shape with peak frequency $\omega_p$ and width $\Delta \omega$ \cite{Schimansky:90,Khovanov:17}. The width is fixed at $\Delta \omega=0.01$, which is close to that of the frequency response of the linearised model (\ref{eqsys}). Three different peak frequencies $\omega_p = \omega_1,\ \omega_2,\ \omega_3$ (as defined in the previous section) are considered. Harmonic noise was simulated by a linear oscillator excited by white noise \cite{Schimansky:90,Khovanov:17}. In the absence of a harmonic signal, the coherence factor $\gamma$ is meaningless, so only the power ratios $e_x$ and $e_v$ were considered.

\begin{figure}[!h]
	\includegraphics[width=2.25in]{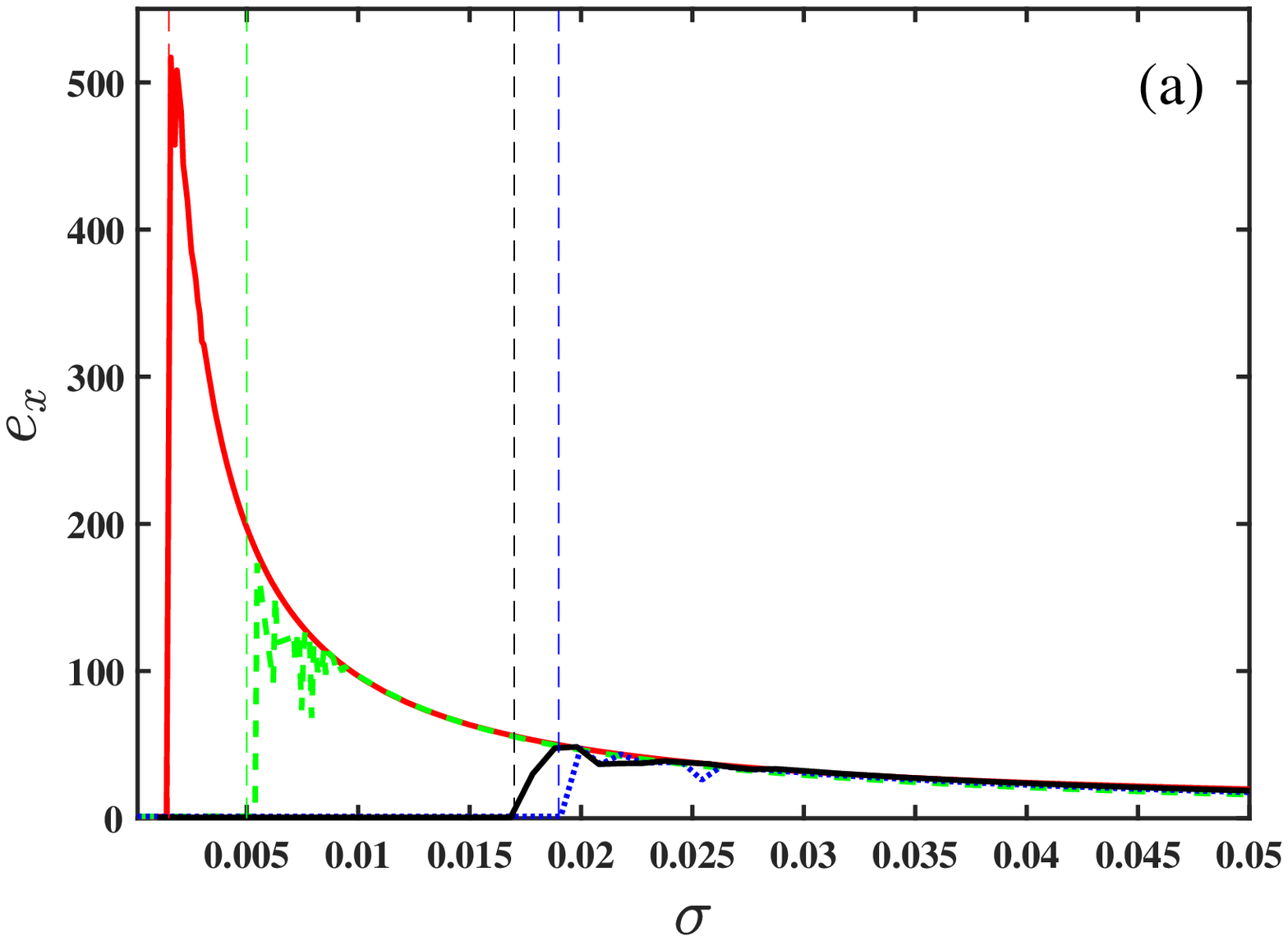}~~\includegraphics[width=2.25in]{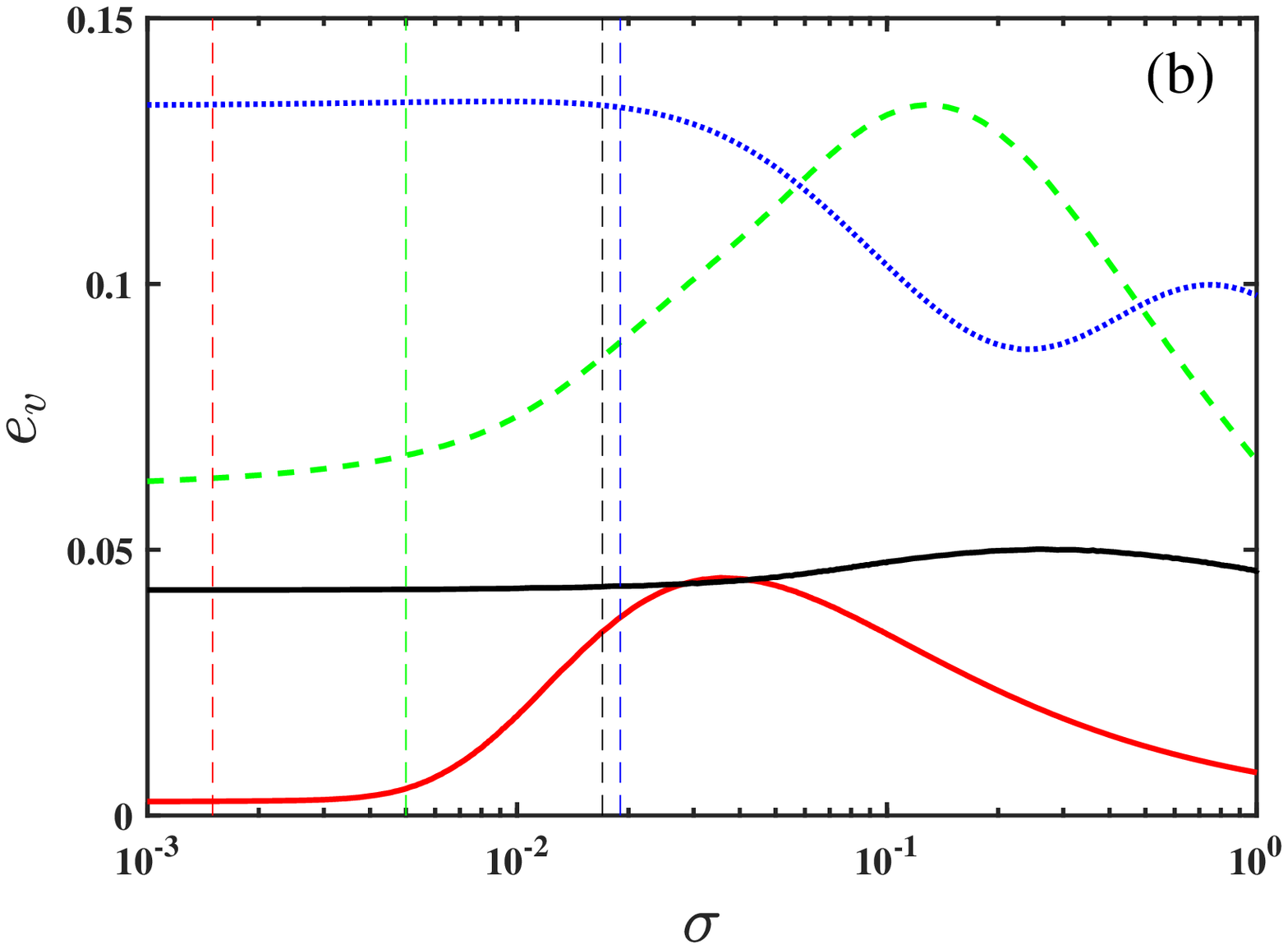}
	\caption{Dependence of power ratios $e_x$ (a) and $e_v$ (b) on noise intensity  $\sigma$. Solid red, dashed green and dotted blue lines correspond to harmonic noise with the peak frequency $\omega_p$ equal to  $\omega_1=0.1$, $\omega_2=0.5$ and $\omega_3=0.9$ respectively. Black solid line shows the ratios for the case of white noise. Vertical dashed lines indicate values of $\sigma$ when the transition to inter-well dynamics occurs.}
	\label{fig2}
\end{figure}

Similarly to the case of the harmonic signal (\ref{eqLFsig}) with frequency $\omega_1$, there is a significant increase in $e_x$ at the point of transition when inter-well dynamics become dominant in (\ref{eqsys}). This increase is observed for both types of noise, and there is an optimal noise intensity, where $e_x$ has a maximum. As noise intensity increases, the distribution of coordinate $x$ changes from a uni-modal shape located around one of the two states to a bi-modal shape that includes both states. The noise intensity $\sigma$ defines the variance of the uni-modal distribution, but in the bi-modal case, the variance is proportional to the distance between the states. The bistability is therefore the leading factor in the sharp rise of $e_x$. In contrast to the case of a harmonic signal, underdamped dynamics do not affect the behaviour of $e_x$ in the case of stochastic excitations. Another difference in the responses to harmonic and stochastic excitations is a striking difference in behaviour of $e_x$ (Fig.~\ref{fig2}(a)) and $e_v$ (Fig.~\ref{fig2}(b)). The maxima of $e_v$ are related to bistability as well, but they are observed at the point when the noise intensity is such that the switching rate between the states is around the relaxation rate of the system. 

For harmonic noise with $\omega_p=\omega_3$, the power ratio $e_v$ has high values at weak noise intensities; this ratio decreases as inter-well dynamics begin to dominate. The existence of a range of $\sigma$ with high ratio $e_v$ illustrates the advantage of a linear design based on condition matching. In this case, the spectrum of harmonic noise matches the frequency response of the system; this leads to efficient harvesting of weak ambient vibrations. In the cases  $\omega_p=\omega_1$ and $\omega_p=\omega_2$, the bistable dynamics of system (\ref{eqsys}) lead to an increase in ratio $e_v$. The influence of $\omega_p$ on the system's response shows that the underdamped dynamics of the harvester are an important factor affecting the efficiency of energy transfer.

\section{Low-frequency harmonic signal with additional signal}

It is believed that the presence of SR and VR causes an additional signal to improve a system's response to a low-frequency harmonic signal. This improvement is observed only when the additional signal is within a narrow range of magnitudes. Since this magnitude affects the mean switching frequency, it has traditionally been assumed that SR and VR manifest when this switching frequency matches that of the low-frequency harmonic signal. In general, however, this condition is not valid, and the phenomena are instead linked to a ``linearisation'' of system dynamics \cite{Dykman:94,Landa:04,Landa:08}, where the amplitude increases enough that the dynamics become effectively monostable. The results of this section illustrate this link.

The low frequency harmonic signal is described by expression (\ref{eqLFsig}). Amplitude $A=0.2$ is used in (\ref{eqgamma}) to estimate the coherence factor $\gamma$ and to calculate the input power $P_{in}=A^2 / 2$ for $e_x$ and $e_v$. Three different forms of additional signal are considered, namely a high-frequency harmonic signal, white noise and harmonic noise. The high-frequency harmonic signal $A_H \sin (\omega_H t)$ with $\omega_H=10$ is used to characterise energy harvesting in a VR regime. White noise of intensity $\sigma$ represents a typical setup for SR. Harmonic noise with $\omega_p=10$ and $\Delta w=0.01$ is used to link SR and VR by considering high-frequency excitation of a stochastic nature instead. Similarly to previous sections, the three frequencies $\omega_1=0.1$, $\omega_2=0.5$ and $\omega_3=0.9$ are considered for the low-frequency harmonic signal (\ref{eqLFsig}).

\subsection{Additional high-frequency harmonic signal}

\begin{figure}[!h]
	\includegraphics[width=2.25in]{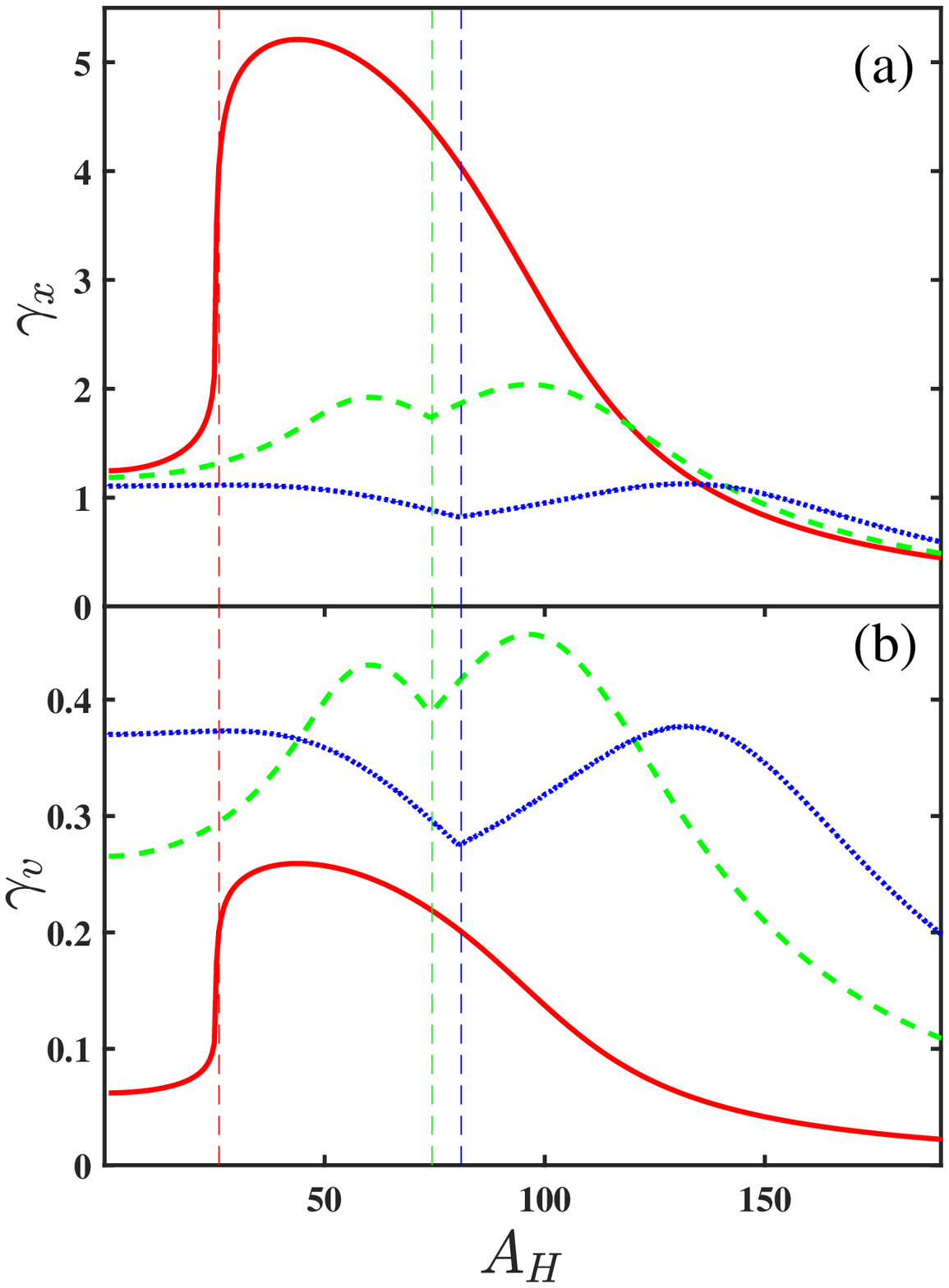}~~\includegraphics[width=2.25in]{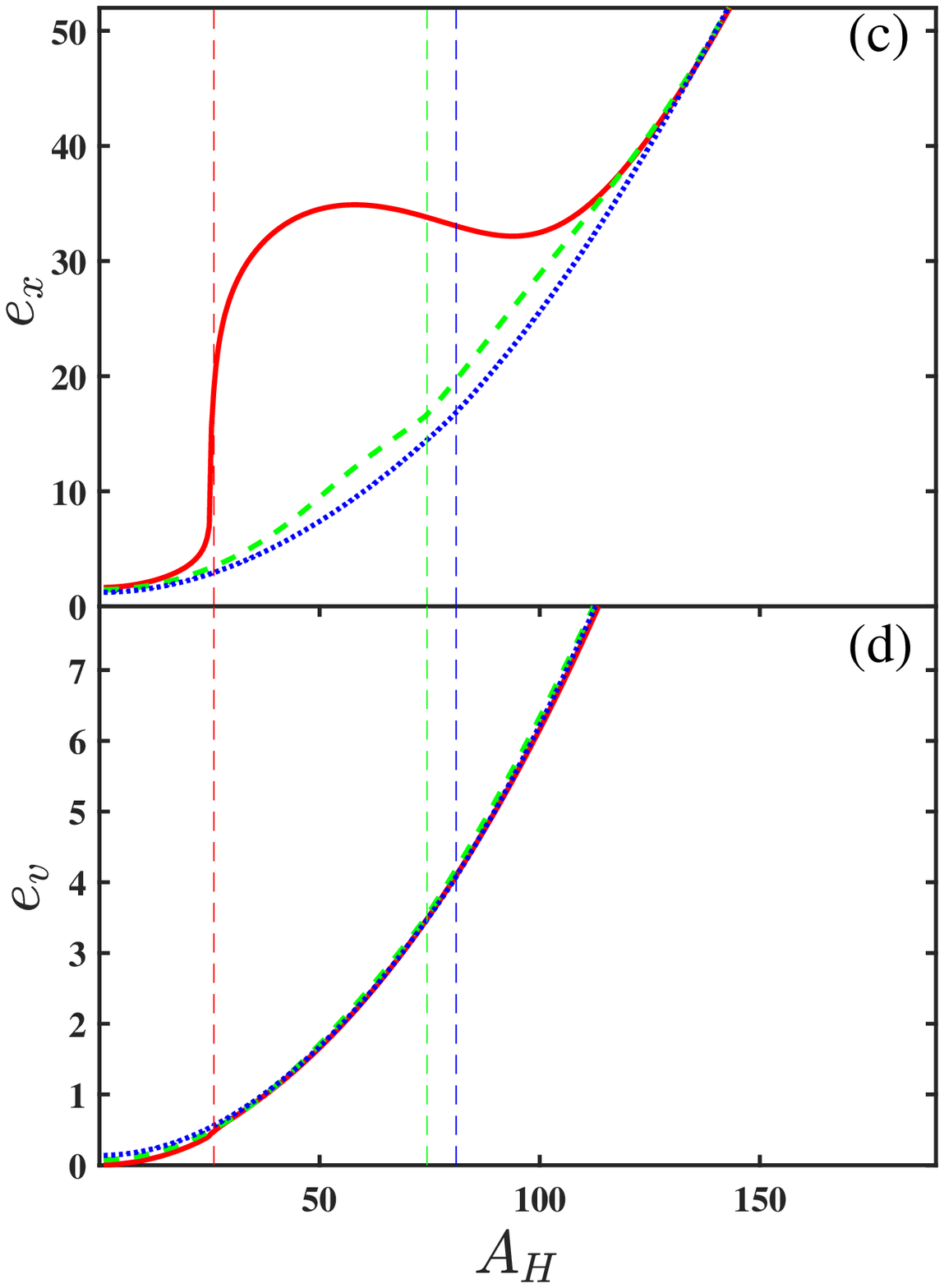}
	\caption{Dependence of coherence factors $\gamma_x$ (a) and $\gamma_v$ (b) and power ratios $e_x$ (c) and $e_v$ (d) on amplitude $A_H$ of the high-frequency harmonic signal. Solid red, dashed green and dotted blue lines correspond to the following frequencies of the low-frequency signal: $\omega_1=0.1$, $\omega_2=0.5$ and $\omega_3=0.9$, respectively. Amplitude of the low-frequency signal is the same for all three frequencies, $A=0.2$. Frequency of the high-frequency component is $\omega_H=10$. Vertical dashed lines indicate values of $A_H$ when the transition to inter-well dynamics occurs.}
	\label{fig3}
\end{figure}

The dependence of the coherence factors on amplitude $A_H$ (Fig.~\ref{fig3}(a, b)) is similar to that shown in Fig.~\ref{fig1}(a, b). An increase in $\gamma$ for $\omega=\omega_1$ is linked to the transition from intra- to inter-well dynamics, whereas this transition leads to local minima for $\omega_2$ and $\omega_3$. The action of the high-frequency harmonic signal is equivalent to an increase in amplitude $A$ or, equivalently, to a reduction in the height of the barrier between the states as $A_H$ increases. The positions of the peaks in $\gamma(A_H)$ depend on the value of $\omega$. For frequencies which are lower than the relaxation rate of the system, the positions are caused by bistability. For frequencies which are close to the natural frequency of the system, the peaks are linked to the system's nonlinearity, and bistability does not play a dominant role. For example, the first peak in $\gamma(A_H)$ for $\omega_2$ is observed in the regime of intra-well dynamics, i.e. dynamics in a monostable potential. The behaviour of ratio $e_x$ (Fig.~\ref{fig3}(c)) illustrates further the difference in the dynamics of the system for different frequencies $\omega$. The ratio has a local maximum in the case of $\omega_1$ only; similarly to the corresponding maximum in $\gamma(A_H)$, this is linked to the transition from  intra- to inter-well dynamics.

Ratio $e_v$ (Fig.~\ref{fig3}(d)) grows monotonically as $A_H$ increases, and the influence of bistability is not pronounced. There is thus no cooperative action between low- and high-frequency signals that could improve the efficiency of energy harvesting in VR regime. 

\subsection{Additional white noise}

\begin{figure}[!h]
	\includegraphics[width=2.25in]{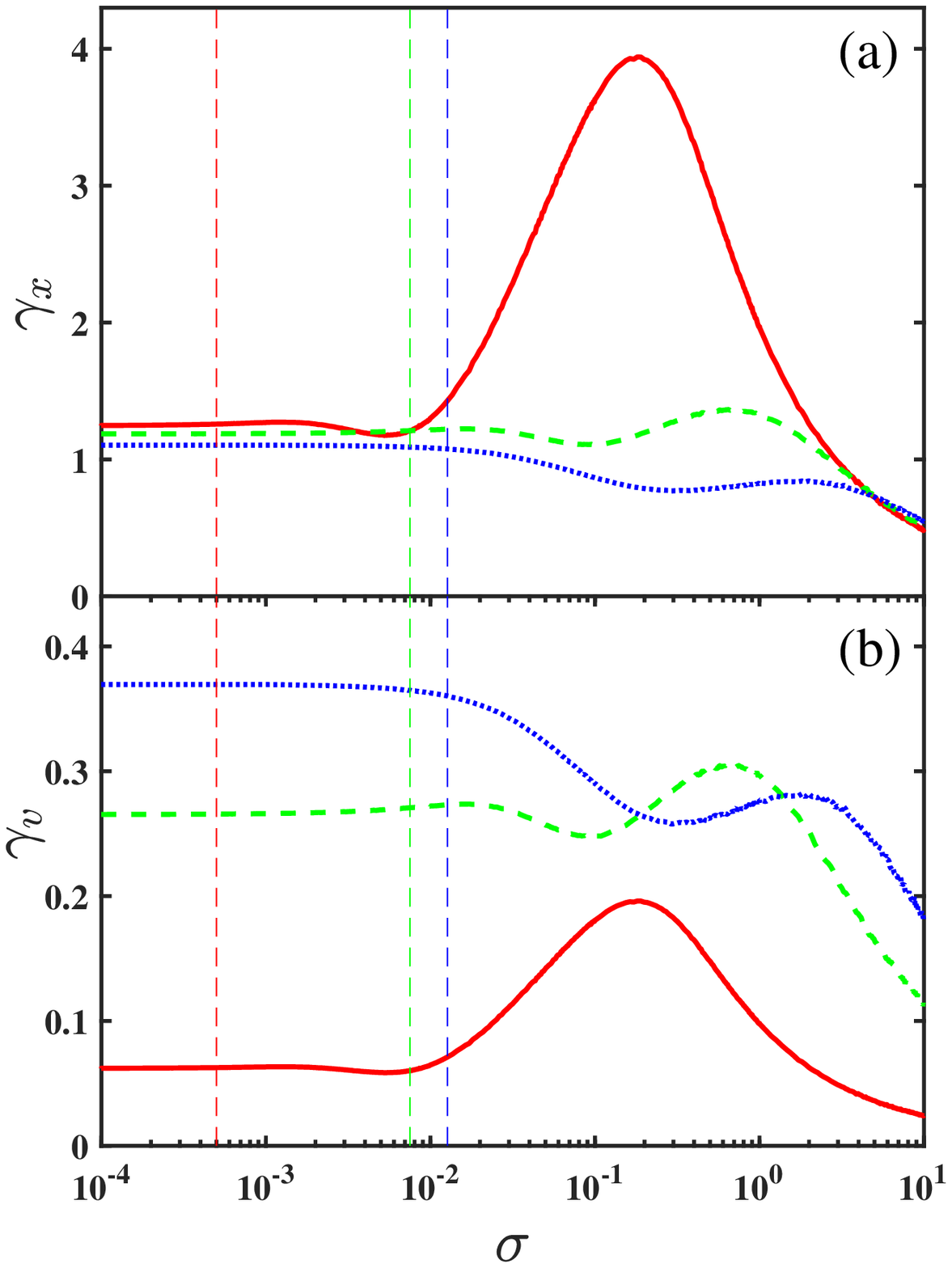}~~\includegraphics[width=2.25in]{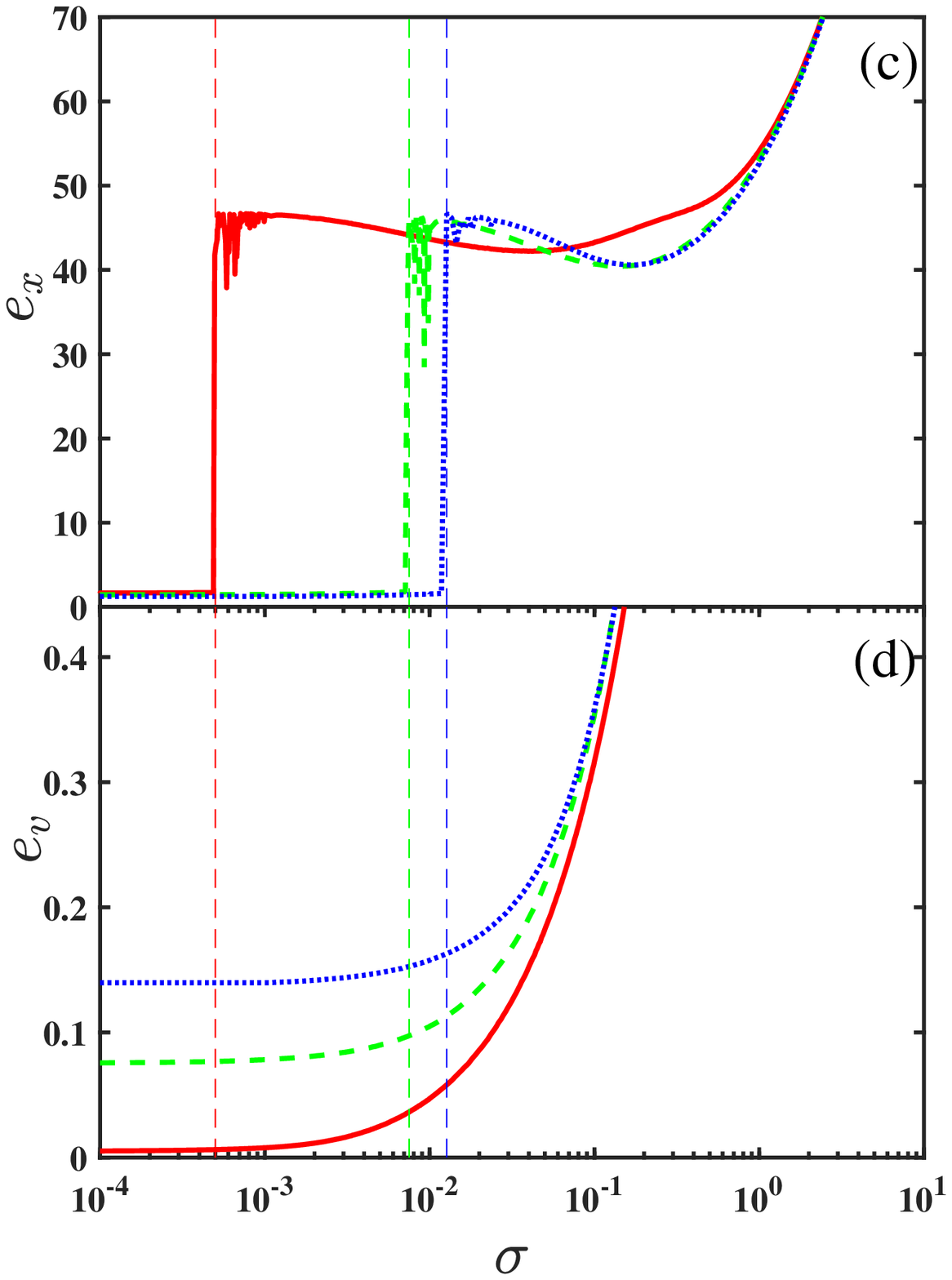}
	\caption{Dependence of coherence factors $\gamma_x$ (a) and $\gamma_v$ (b) and power ratios $e_x$ (c) and $e_v$ (d) on intensity $\sigma$ of white noise. Solid red, dashed green and dotted blue lines correspond to the following frequencies of the low-frequency signal: $\omega_1=0.1$, $\omega_2=0.5$ and $\omega_3=0.9$, respectively.  Amplitude of the low-frequency signal is the same for all three frequencies, $A=0.2$. Vertical dashed lines indicate values of $\sigma$ when the transition to inter-well dynamics occurs.}
	\label{fig4}
\end{figure}

The low-frequency signal (\ref{eqLFsig}) excites periodic oscillations within one of the wells of bistable system (\ref{eqsys}). The height of the barrier between the two states depends on the frequency $\omega$; the barrier is lowest for $\omega_1$ and highest for $\omega_3$. The location of peaks in Fig.~\ref{fig4}(a, b) follows the same pattern: $\sigma$ is smallest for $\omega_1$ and largest for $\omega_3$. The peaks represent the SR phenomenon and correspond to the point where the barrier height is close to the noise intensity, i.e. where the dynamics become effectively monostable. 

For coordinate $x$, the energy transfer from noise becomes more efficient when switching between states begins (Fig.~\ref{fig4}(c)). This result is similar to those presented in Fig.~\ref{fig2}(a). Bistability leads to a sharp increase in ratio $e_x$ for all three values of frequency $\omega$. The distance between the two states defines the magnitude of this step-like increase. The monotonic behaviour of ratio $e_v$ shows that the SR phenomenon also does not contribute to the efficiency of energy harvesting.

\begin{figure}[!h]
\centering \includegraphics[width=2.5in]{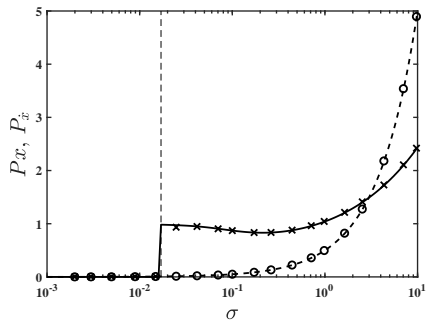}
	\caption{Dependence of variances of coordinate $x$, $P_x$, and velocity $\dot x$, $P_{\dot{x}}$,  on intensity $\sigma$ of white noise. Lines show variances obtained using Eq. (\ref{Browndist}) and by the assumption that $\psi=0$. Markers show variances calculated numerically by simulating the system (\ref{eqsys}). Note that the same simulation was used in Fig.~\ref{fig3} (black solid lines). Solid line and marker $\times$ correspond to $P_x$, and dashed line and marker $\circ$ correspond to $P_{\dot{x}}$.  Vertical dashed line indicates a value of $\sigma$ when the transition to inter-well dynamics occurs.}
	\label{fig6}
\end{figure}

The observed difference in the dependence of the ratios $e_x$ and $e_v$ on the noise intensity $\sigma$ is defined by the behaviour of the variances of $x$ and $v$, respectively. For both ratios, the denominator is the input power $P_{in}=A^2/2$, and it is the same. However, the numerator in the ratios is the output powers $P_x$ or $P_v$, which as it followed from Eqs. (\ref{eqPx}) and (\ref{eqPv}) are the variances of $x$ and $v$ respectively. Since the back action of the piezoelectric on the beam is small, the output power $P_v$ could be assumed to be  directly proportional to the variance of velocity $\dot{x}$, $P_{\dot{x}}$. Let us ignore the back action by assuming that parameter $\psi$ in  (\ref{eqsys}) is equal to zero. Then,  the first equation in  (\ref{eqsys}) is decoupled from the second equation. In the case of excitation $F(t)$ in the form of  white noise, the first equation is described Brownian motion in a potential. The joint equilibrium  distribution for the coordinate and velocity for Brownian motion is well-known (see, for example, \cite{Schimansky:93}):
\begin{align}\label{Browndist}
	\rho(x,\dot{x})=\frac{1}{Z} \exp\left( -\frac{\alpha}{\sigma} (\dot{x})^2\right) \exp\left( -\frac{ 2 \alpha }{\sigma} U(x) \right)  \ .
\end{align}
Here the potential $U(x)=-0.25 x^2+ 0.125 x^4$ corresponds to the nonlinearity in the system (\ref{eqsys}) and $Z$ is a normalising factor. This distribution is defined by two separate exponential factors, each of which depends on either velocity $\dot{x}$ or coordinate $x$. Expression (\ref{Browndist}) shows that the variance of velocity $P_{\dot{x}}$ does not depend on the system nonlinearity (the potential $U(x)$) and the variance is proportional to noise intensity $\sigma$. In contrast, the variance of coordinate $P_x$ is defined by both noise intensity $\sigma$ and the shape of $U(x)$.  Both variances obtained using the distribution  (\ref{Browndist}) show behaviour (Fig.~\ref{fig6}) which is similar to those for the power factors in Fig.~\ref{fig4} (c) and (d). Quantities calculated for coordinate $x$ ($e_x$ in Fig.~\ref{fig4} (c) and $P_x$ in Fig.~\ref{fig6}) demonstrate a sharp increase at the transition point from intra- to inter-well dynamics. In contrast, $e_v$ in Fig.~\ref{fig4} (d) and $P_{\dot{x}}$ in Fig.~\ref{fig6} show the monotonic increase with the growth of noise intensity $\sigma$. These results illustrate that the nonlinearity of coordinate $x$ and the linear damping term for velocity $\dot{x}$ defines the non-monotonic and monotonic dependence of the power ratios $e_x$ and $e_v$ respectively.

Fig.~\ref{fig6} includes also the results of numerical simulations of the system (\ref{eqsys}) with non-zero parameter $\psi$. The variances $P_x$ and $P_{\dot{x}}$ calculated theoretically and numerically show an excellent correspondence. However, this observation does not mean that $P_{\dot{x}}$ can be used instead $P_v$ for an estimation of the power ratio $e_v$. The variance $P_v$ is defined by the spectral content of velocity $\dot{x}$ and the frequency response of the second equation of  system (\ref{eqsys}). As a result, the link between $P_v$ and $P_{\dot{x}}$ is more complicated than direct proportionality. This conclusion is confirmed by a non-monotonic behaviour of ratio $e_v$ in the case of excitation by white noise only (black curve in Fig.~\ref{fig3}(b)). Note, that in this case, the dependence of $P_{\dot{x}}$ on noise intensity $\sigma$ is close to a horizontal line.

\subsection{Additional harmonic noise}

\begin{figure}[!h]
\includegraphics[width=2.25in]{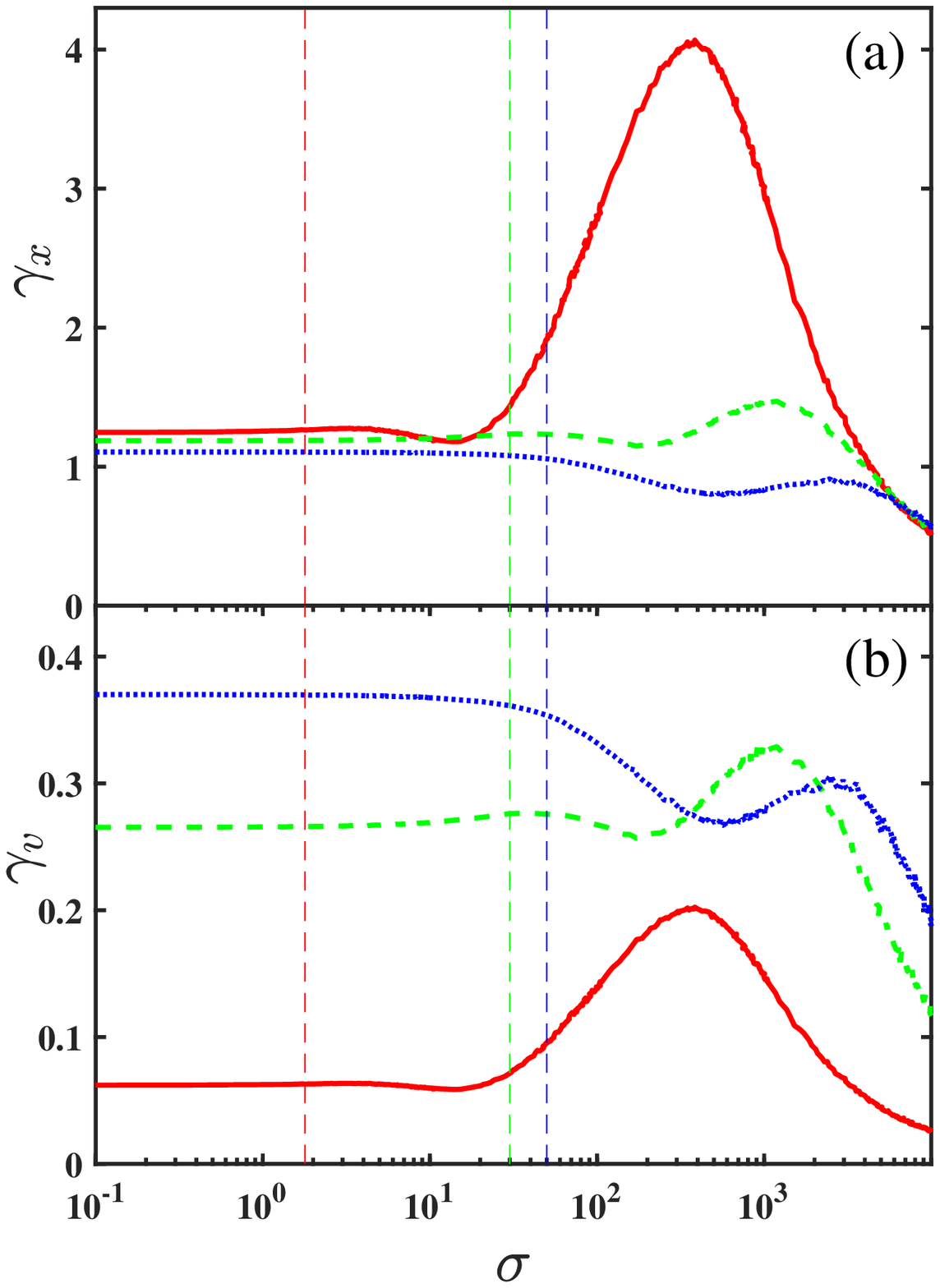}~~\includegraphics[width=2.25in]{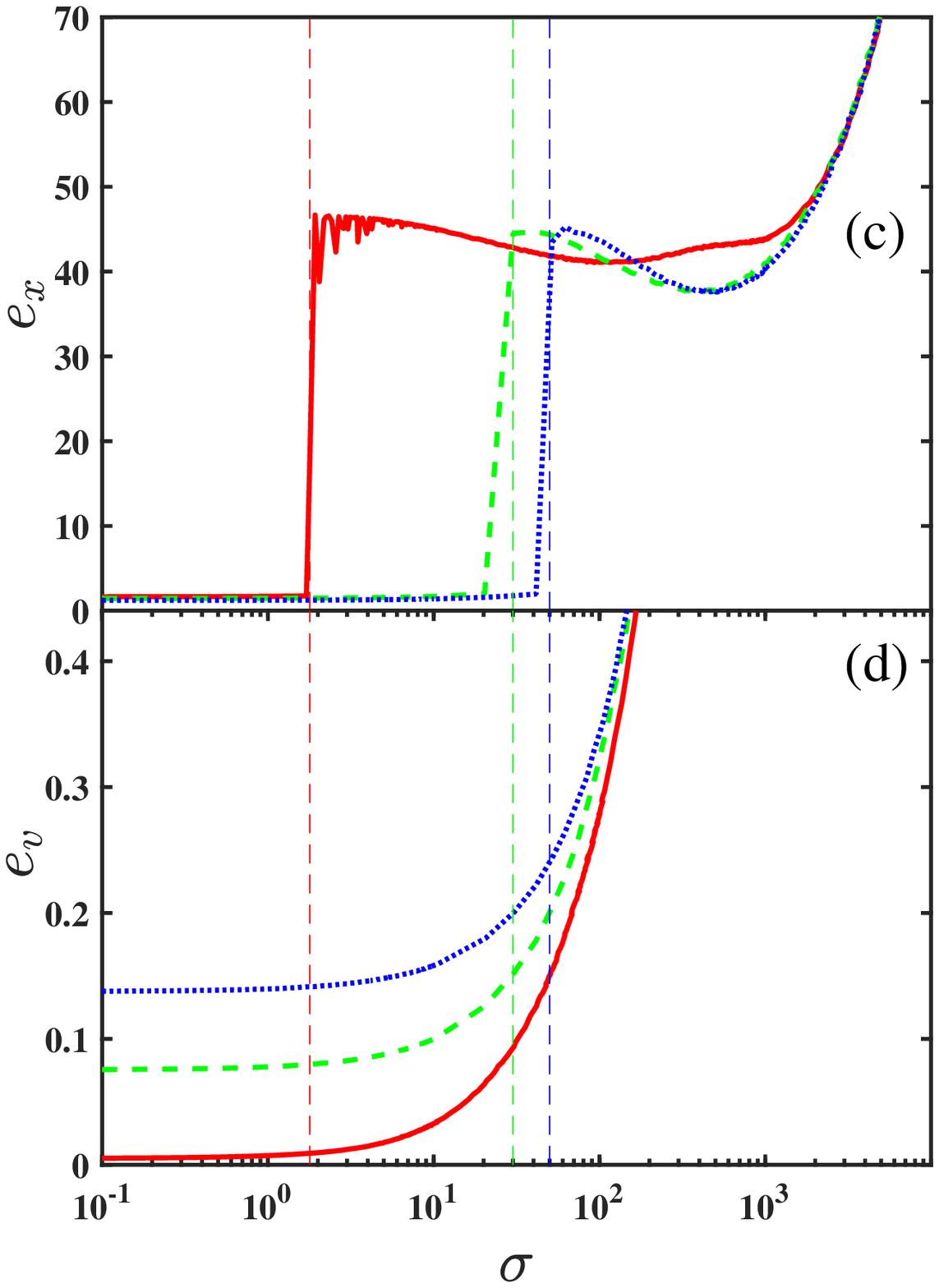}
\caption{Dependence of coherence factors $\gamma_x$ (a) and $\gamma_v$ (b) and power ratios $e_x$ (c) and $e_v$ (d) on intensity $\sigma$ of harmonic noise. Solid red, dashed green and dotted blue lines correspond to the following frequencies of the low-frequency signal: $\omega_1=0.1$, $\omega_2=0.5$ and $\omega_3=0.9$, respectively. Amplitude of the low-frequency signal is the same for all three frequencies, $A=0.2$. Harmonic noise has peak frequency $\omega_p=10$ and spectral width $\Delta \omega=0.01$. Vertical dashed lines indicate values of $\sigma$ when the transition to inter-well dynamics occurs.}
\label{fig5}
\end{figure}

The relationships shown in Fig.~\ref{fig5} for high-frequency harmonic noise are nearly identical to those obtained for white noise (Fig.~\ref{fig4}); the only difference is the magnitude of noise intensity $\sigma$. This similarity demonstrates that system (\ref{eqsys}) strongly suppresses high-frequency components of external excitations, and that it is the wide tails of the (Gaussian) distribution of the stochastic excitations that determine the system response. In contrast, the distribution of a high-frequency harmonic signal is bounded. SR and VR therefore manifest differently. For a high-frequency harmonic signal, the coherence factor $\gamma$ and power ratio $e$ depend on the underdamped dynamics (i.e on frequency $\omega$), and peaks in $\gamma$ and $e$ are determined both by nonlinearity in the vicinity of each state and by bistability. A stochastic signal ''smooths'' the dynamics around each state, and the influence of intra-well nonlinearity is weak. Bistability is a thus dominant factor in the system response to a stochastic signal.

\section{Conclusion}

These results show that bistability can lead to an improvement in harvester efficiency. However, the system must be tuned using properties of the external excitations. The underdamped dynamics of the harvester are a significant factor affecting both the system response and manifestations of SR and VR. If the harmonic signal frequency is less than the relaxation rate of the system, then the point of transition from intra- to inter-well dynamics determines the excitation magnitude that peaks in coherence factors appear at. However, this transition instead leads to a minimum in $\gamma$ if the frequency of the harmonic signal is close to a resonance frequency of the system. The influence of underdamped dynamics on SR is not as dramatic. However, the peak in coherence factors can appear in both monostable and bistable regimes, depending on the frequency of the harmonic signal. 
Note that smaller damping leads to the dominance of chaotic dynamics near the point of transition from intra- to inter-well dynamics.

Both SR and VR have no significant influence on harvester efficiency since it is the velocity $v$ of the mechanical part that determines the energy transfer. In contrast, coordinate $x$ undergoes bistability in the harvester. Due to this, the increase in power ratio $e_x$ is strongly linked to bistability. Therefore, in order to effectively utilise a bistable design, the dynamics of the velocity should also demonstrate bistability. Such a transformation could be achieved, for example, by using additional piezoelectric components to feed back some of the harvested electrical energy.

A comparative analysis of the system response to harmonic noise and a harmonic signal shows that the main difference between SR and VR is in the distributions of the signals acting on the system. The distribution has wide tails in the SR case, but it is bounded in the case of VR.  The manifestation of SR is similar for white and harmonic noise. However, the system response to a single input in the form of harmonic noise depends on the location of the peak frequency $\omega_p$. Inter-well dynamics are observed for low values of $\omega_p$, but the values of power ratio $e_v$ are higher for frequencies $\omega_p$ which are close to the resonance frequencies of the system.

\end{document}